# Complex structures in the Au – Cd alloy system: Hume-Rothery mechanism as origin

**Valentina F Degtyareva and Nataliya S Afonikova**

Institute of Solid State Physics, Russian Academy of Sciences, Chernogolovka 142432, Russia

E-mail: degtyar@issp.ac.ru

**Abstract.** The binary (simple metal) phase diagram Au–Cd contains a number of intermetallic compounds with various distortions, superlattices and vacancies. To understand the reasons of these structural complexities and their phase stability, we analyze these crystal structures within the nearly free-electron model in the frame of Fermi sphere – Brillouin zone interactions. Examination of the Brillouin-Jones configuration in relation to the nearly-free electron Fermi sphere provides insights for significance of the valence electron energy contribution to the phase stability. Representation of these complex structures in the reciprocal space clarifies their relationship to simple basic cells. This approach shows the importance of the additional planes for the stability of superslattices. The AuCd-$hP$18, AuCd$_3$-$hP$24 and AuCd$_4$-$hP$273 structures are shown to be related to the AuCd-$cP$2 via rhombohedral distortion with superlattices and vacancies.

## 1. Introduction

The binary alloy system Cu – Zn represents a classical example of Hume-Rothery phases that include the key metallic close-packed structures: face-centered cubic (*fcc*), hexagonal close-packed (*hcp*) and body-centered cubic (*bcc*). The sequence of the phases as a function of the average number of valence electrons per atom *fcc* → *bcc* → complex γ-phase → *hcp* is defined by the following values of the valence electron concentration 1.35 →1.5 → 1.62 → 1.75 (electron/atom).

The fact that the phase boundaries are defined by these numbers of the valence electron concentration has been explained within the free-electron model that takes into account the contact of the Brillouin zone (BZ) planes and the Fermi sphere (FS). This contact results in a formation of the energy gap and significant gain in the band structure energy term related to the valence electron energy [2,3]. Similar phase sequences were observed in many other binary alloy systems containing a noble metal from group IB and a neighboring element from groups IIB – VB. The Cu–Zn alloy system is relatively simple system with only four intermediate phases that exhibit quite simple structures with the ground high-symmetry metallic phases like *fcc*, *bcc* and *hcp*. Along with these basic metallic structures there are phases Cu$_5$Zn$_8$-$cI$52 and CuZn$_3$-$hP$3 that are related to *bcc* through superlattices, distortions and vacancies [4]. The crystal structures of binary phases were found to represent complex distortive variants of the high-symmetry structures. Examples of binary phases derivatives from *bcc* and *hcp* through the formation of long-period superlattices, distortions and vacancies were considered recently [4,5].

In this paper we consider intermediate phases in the Au–Cd alloy system that is isoelectronic to the Cu–Zn system. Atomic size and electronegativity differences for Au and Cd are nearly same as for Cu and Zn, however the phase diagram for Au–Cd is much more complicated compared to Cu–Zn [6,7]. Formation of structurally complex Au–Cd phases may have its origin in the different contribution of two main components of the crystal structure energy: the electrostatic and the band structure terms. The former prefers high-symmetry close-packed structures and the latter defines structural complexity.

Compared to lighter elements of the same column in the periodic table, heavier elements are characterized by more complex electronic core shells which change the balance between the crystal structure energy terms. Ionic interaction (the electrostatic term) is reduced and the band structure term therefore may increase leading to significant differences in structural behavior. Gold is considered as an element with unique properties that can be accounted by "a stronger screening of electron orbitals" and "an increase of the effects on valence-shell properties down a column of the periodic table" [8].

Comparing phase diagrams of the isoelectronic systems Cu–Zn, Cu–Cd and Au–Cd one can see that the few and relatively simple phases in Cu–Zn change to structurally complex compounds in Cu–Cd and to a variety complex phases in Au–Cd. According to the data from Pearson [9], atomic ratio Cd/Cu is 1.227 that is essentially larger than atomic ratios Zn/Cu and Cd/Au that are equal to ~1.09. The binary alloy system Cu–Cd that contains four intermetallic compounds $Cu_2Cd$, $Cu_4Cd_3$, $Cu_5Cd_8$, and $Cu_3Cd_{10}$ with 12, 1124, 52 and 28 atomic sites per unit cell has been recently analyzed with the help of the Hume-Rothery rules and the Jones model to account for the factors controlling structure stability [10]. Formation of complex structures in the Cu–Cd system is believed to be influenced by the atomic size difference.

Intermetallic phases in the Au–Cd system prefer long-period ordered superstructures over the simple high-symmetry metallic structures that are observed in Cu–Zn. As shown in this paper, one of the reasons for the formation of the superstructures is that for Au and Cd atoms the difference in atomic numbers (79 and 48, respectively) provides relatively high structure factor for the superstructure reflections due to atomic site ordering. The energy gap is deeper on the BZ planes with stronger structure factors which stimulates stabilization of complex site ordered structures among Au–Cd alloys.

## 2. Theoretical background and method of analysis

The Hume-Rothery mechanism supposes a formation of the energy gap due to the contact of a Brillouin zone plane and the free-electron Fermi sphere. The symmetrical close-packed structures *fcc*, *bcc* and *hcp*, together with the complex cubic γ-phase $Cu_5Zn_8$-*cI*52, form the basic Hume-Rothery phase sequence. In binary alloys there is a vast variety of structures arising due to the importance of some other factors such as the difference in atomic sizes, electronegativity etc. Beyond these factors, formation of stoihiometric compounds at certain alloy composition is defined by effects of the FS-BZ interaction. The Hume-Rothery mechanism has been identified to play a role in the stability of structurally complex alloy phases, quasicrystals and their approximants [10-13]. Jones model can be used to account for phase stability in tetrahedral cluster structures, icosahedral and trigonal-prismatic clusters as building blocks. Formation of the complex structures of elemental metals under pressure can also be related to the Hume-Rothery mechanism [14-17].

The crystal energy contribution due to the band structure energy can be estimated by analysing configurations of Brillouin-Jones zone planes in the nearly free-electron model. A special program BRIZ has been developed [18] to construct FS-BZ configurations and to estimate some parameters such as the Fermi sphere radius ($k_F$), values of reciprocal wave vectors of BZ planes ($q_{hkl}$) and volumes of BZ and FS. The BZ planes are selected to match the condition $q_{hkl} \approx 2k_F$ that have a remarkable structure factor. In this case an energy gap is opened on the BZ plane leading to the lowering of the electron band energy. The ratio of ½ $q_{hkl}$ to $k_F$ is usually less than 1 and equals ~0.95; it is called a "truncation" factor [19]. In the FS-BZ presentations by the BRIZ program the BZ planes cross the FS, whereas in the real system the Fermi sphere is deformed and accommodated inside BZ. The "truncation" factor has a characteristic value and corresponds to a decrease in the electron energy on the BZ plane.

The crystal structure of a phase chosen for the analysis by the BRIZ program is characterized by the lattice parameters and the number of atoms in the unit cell, which define the average atomic volume ($V_{at}$). The valence electron concentration (z) is the average number of valence electrons per atom that gives the value of the Fermi sphere radius $k_F = (3\pi^2 z / V_{at})^{1/3}$. Further structure characterization parameters are the number of BZ planes that are in contact with the FS, the degree of "truncation" factor and the value of BZ filling by electronic states, defined as a ratio of the volumes of FS and BZ. It should be noted that in the current model estimation of the FS volume is a measure of the number of valence electrons participating in the band structure contribution although in the real case the FS is deformed from a sphere and adopted to the BZ.

Presentations of the FS-BZ configurations are given with the orthogonal axes with the following directions in the common view: a* is looking forward, b* - to the right and c* - upward. For the hexagonal lattice in the reciprocal space a*= $a_{1h}$*cos30º, b*= $a_{2h}$* and c*= $c_h$*. Structural data for binary phases considered in this paper have been found in the Pauling File [7] and in recent papers cited in the corresponding sections of this paper.

## 3. Results and discussion

The phase diagram Au–Cd is presented in Fig. 1 following Ref. [6]. For all known intermediate Au–Cd phases, we illustrate Brillouin-Jones zones constructed with the program BRIZ. Structural data for the phases and data obtained with the BRIZ program are presented in Tables 1 and 2. Table 1 includes structure types discussed in previous papers [4,5]. Table 2 presents structures that are discussed in this paper while the diffraction patterns and BZ-FS configurations for these structures are shown in Fig. 2.

*3.1. The gold-rich alloys of the Au-Cd system*

The Au(Cd) solid solution extends to about 33 at% Cd. Thus the *fcc* boundary satisfies the Hume-Rothery rule ~1.36 electron/atom, as for *fcc* Cu(Zn) phase. At low temperature several long-period superlattices were observed [6,7]. We consider here three representative supercells based on the close-packed *fcc* and *hcp* structures.

*3.1.1. Au$_3$Cd-tI16 phase*
The long-period phase derived from *fcc* is formed at ~25 at% Cd by cooling from the *fcc* Au(Cd) solid solution. The Au$_3$Cd-*tI*16 structure contains nearly four *fcc* cells along the *c* axis with additional site ordering of components [20]. The ordered alloy Au$_3$Cd has one-dimensional antiphase domain structure. Basic face-central tetragonal cell contains the step shift by a translation vector (**a$_1$**+**a$_2$**)/2. A key factor for stability of this structure is arising of the new diffraction peak (105) and the appearance of the corresponding BZ plane at the FS (see Table 2 and Fig. 2a). Similar structure with *fcc* superlattices are also known for Ag$_3$Mg and Pt$_3$Sb [7].

*3.1.2. Au$_2$Cd-hP4 phase*
Alloys with 25 – 29 at% Cd were found to crystallize in the hexagonal close-packed cell with the *c*-axis twice as long as the basic *hcp* cell [20]. This structure is usually called as double hexagonal close-packed, *dhcp*, Pearson symbol *hP*4. Similar structure was found in other noble-metal based alloys at compositions ~1.25 – 1.30 electron/atom. At these values of electron concentration there is a special condition where FS is touched by the edges of crossing planes (100) and (101). This brings an additional reduction in the band structure energy (see Table 1 and Fig. 1).

An example of this kind of superlattice was discussed for Au$_{87}$In$_{13}$ -*hP*4 phase [5]. One more example is a more complex superlattice of *hcp* in the alloys Au-10 at% Sn with *hP*16 structure [7]. Lattice parameters of this phase are related to the basic *hcp* as $a = 2a_0$ and $c = 2c_0$ with the axial ratio $c_0/a_0 = 1.642$, same as for the basic cell of Au$_2$Cd-*hP*4 phase. This $c_0/a_0$ ratio is slightly higher than the ideal value $c/a = \sqrt{8/3} = 1.633$, because for z = 1.3 electron/atom, $k_F$ is less than the distance to the (002) plane. With this configuration one should expect an attraction of FS to the (002) planes, that should produce an increase of *c/a* comparing with the ideal value.

*3.1.3. Long-period superlattice Au₂Cd-hP98 (29-34 at% Cd)*

Along with *hcp*-based superlattices that have long period in *c* dimension such as in *hP*4, there are also several phases with long periods in *a* dimension. We consider the phase that is formed at 29 – 34 at% Cd and described in Refs. [21,22] as the basic *hcp* cell with $a = 7a_0$ and $c = c_0$ (see Table 1 and Fig. 1). An additional factor of stability for this *hP*98 phase appears as evident from the configuration of the BZ in a form of a decagonal-based prism inscribed in the FS in comparison to the hexagonal prism for the usual *hcp*, as was discussed in [5].

*3.2. Au-Cd alloys with 43 – 57 at% Cd*

The high-temperature phase has cubic *cP*2 structure of CsCl type, as for CuZn low-temperature phase. The Hume-Rothery rule for stability of *bcc* at ~1.5 el/atom is satisfied. At lower temperatures AuCd-*cP*2 phase undergoes martensitic transitions with formation of several different distorted structures and superlattices depending on composition [6,7].

*3.2.1. Martensitic AuCd phases mP6 and oP4*

Distorted structures related to *bcc* were considered with analysis of the FS-BZ configuration in Ref. [4]. Lattice distortions and superlattice formation due to transitions from cubic (*cP*2) to monoclinic (*mP*6) [23] or orthorhombic (*oP*4) [24] lead to the appearance of new BZ planes close to the FS and, as a consequence, to the gain in the band structure energy (see Table 1 and Fig. 1).

*3.2.2. The superstructure AuCd -hP18*

Close to the stoihiometric composition 50/50, the AuCd-*hP*18 phase has been observed [25-27]. The high-temperature cubic AuCd phase transforms to *hP*18 through a rhombohedral distortion with atomic displacements and a superlattice formation. The basic hexagonal cell can be considered as *hP*3 with lattice parameters $a_0$ and $c_0$ that related to AuCd-*hP*18 as $a = a_0\sqrt{3}$ and $c = 2c_0$. This transformation is related to the known transition in Ti-alloys: $bcc – hP3$ (or $bcc – \omega$-phase) [4]. With this structural transformation the number of BZ planes in the vicinity to FS increased: instead of 12 planes of the (110) type for AuCd-*cP*2, there are 18 planes for *hP*3 and 30 planes for *hP*18. Therefore the degree of BZ filling by electron states increases from 0.77 to 0.90, respectively (see Table 2 and Fig. 2b). Space group for AuCd-*hP*18 was suggested as (143) *P*3 and later rejected in favor of (147) *P*-3 [27]. The latter space group is used for our BZ constructions.

Similar transformation from cubic *cP*2 to a trigonal structure was observed for AgCd compound and explained as the shape memory effect driven the by Fermi-surface mechanism [28].

*3.3. Alloy phases in region 61 – 67% Cd*

*3.3.1. The Au₅Cd₈-cI52 phase*

The classical gamma-brass phase $Cu_5Zn_8$ is assumed to correspond to the electron concentration 21/13 = 1.615. Alloys Au–Cd display several phase regions with the temperature variation: $\delta \rightarrow \delta' \rightarrow \delta''$ and the separate region $Au_3Cd_5$ [29]. The high temperature phases δ and δ´ were found to have structures similar to the $Cu_5Zn_8$-*cI*52, satisfying the Hume-Rothery rules (see Table 2 and Fig. 2c).

*3.3.2. The Au₃Cd₅-tI32 phase*

This phase is stable at room temperature, and corresponds to a specific value of composition of the $Au_3Cd_5$ [29]. Interestingly, that phase regions on the T – c diagram for $Au_3Cd_5$-*tI*32 and $Au_5Cd_8$-*cI*52 are situated close in composition; however the *tI*32 region exists separately (see Fig. 1). This corresponds to absence of structural relationship for these two types of structures. Atomic arrangements for the $Au_3Cd_5$-*tI*32 structure relate to the tetrahedral packing of ordered atoms due to their slight atomic size difference.

Energetic preference for Au$_3$Cd$_5$-$tI$32 over Au$_5$Cd$_8$-$cI$52 can be accounted by a better V$_{ES}$/V$_{BZ}$ filling which is equal to 95% and 93%, respectively (see Table 2 and Fig. 2d).

According to the symmetry of space group *I*4/*mcm* in one *ab* layer ⊥*c* at z = 0 Au atoms in position 8h form square-triangle nets and Cd atoms in position 16k form octagon-square nets. Interestingly, these two types of atomic nets are related to the arrangements of atoms in recently found incommensurate elemental structures under high pressure, as discussed for potassium [30] and calcium [31,32].

*3.3.3. Complex phase AuCd$_2$ -mC72*

Recent structural studies of AuCd$_2$ alloys (high-temperature form δ″) have determined the incommensurately modulated structure described with the average cell *mC*72 [33]. The basic structure of δ″-AuCd$_2$ represents a new type; the elemental ordering causes an incommensurate modulation. Structural fragments of the δ″-phase resemble that of γ- brass, but the cell content of the δ″-phase (72 atoms) does not correspond to any simple multiple of the primitive cell of γ-brass (26 atoms). Construction of the Brillouin zone with planes lying close to the FS resulted in a many-faced polyhedron that is 89% filled by electron states, which corresponds to the Hume-Rothery mechanism of the structural stabilization (see Table 2 and Fig. 2e).

*3.4. The AuCd$_3$ -hP24 phase (72 – 76% Cd)*

At the alloy composition AuCd$_3$ (z = 1.75), a phase has been found with the structure *hP*24 [34]. This structure resembles the CuZn$_3$-*hP*3 phase known as high-temperature phase in the Cu-Zn alloys at the same valence electron concentration z =1.75 [4]. Lattice parameters of AuCd$_2$ -*hP*24 are related to that of the basic cell as $a = a_0\sqrt{3}$ and $c = 3c_0$. The number of basic cells is 9 and $c_0/a_0 = 0.603$. There are some vacancies present in this structure, because instead of 27 atoms per cell there are 24 atoms per cell and with the ratio 24/9 one obtains 2.67 atoms per basic cell. The CuZn$_3$-*hP*3 phase is characterized by the presence of vacancies as well and has ~2.7 atoms/cell. Axial ratio $c_0/a_0$ is comparable with the $c_0/a_0 = 0.612$ of the ideal *hP*3 cell formed from *bcc*. Construction of the FS-BZ configuration for AuCd$_3$ -*hP*24 one obtains additional planes arising due to the supercell formation (see Table 2 and Fig. 2f).

There is a relation between AuCd$_3$ -*hP*24 and the AuCd-*hP*18 structure considered above in 3.2.2. This is because both phases are superlattices to the ω-Ti type *hP*3 structure with supercells having $a = a_0\sqrt{3}$ and $c = 2c_0$ in the case of *hP*18 and $c = 3c_0$ in the case of *hP*24. Basic cells have $c_0/a_0$ close to the ideal value ~ 0.612 and the number of atoms is 3 and 2.67, respectively. The number of atoms is reduced due to the increase in the valence electron concentration keeping the value of zN nearly constant (3×1.5 = 4.5 and 2.67×1.75 = 4.67, respectively). By using only very strong reflections we obtain BZ similar to a basic simple structure of *hP*3-type assuming relations of (300) and (113) to (110) and (101) of *hP*3 (Fig.2f, right).

Another phase reported for AuCd$_3$ is *tP*2 [35]. These studies need confirmation. FS-BZ configurations for this structure is shown on Fig. 1, data are given in Table 1.

*3.5. The AuCd$_4$ -hP273 (82 at.% Cd)*

Crystal structure of a phase with the composition Au-80 at.% Cd was determined very recently [36]. Structure of AuCd$_4$-*hP*273 is closely related to AgMg$_4$ with the space group P6$_3$/m and the number of atoms ~92 [37]. The compound AuCd$_4$ represents a superstructure of the AgMg$_4$ type with parameters √3a × √3a × c; the unit cell contains ~273 atoms. Both compounds are characterized by valence electron count ~1.8 and merited as the Hume-Rothery phases.

The group of diffraction peaks close to 2k$_F$ was selected for construction of BZ (see Table 2 and Fig. 2g). For AuCd$_4$-*hP*273 constructed BZ has highly spherical form, accommodating well the FS with filling ~96%. Considering only the strongest reflections it is possible to trace structural relationship of hP273 to a basic simple structure of hP3-type assuming relations of (630) and (145) BZ planes to (110) and (101) of *hP*3 (Fig.2g, right). The cell of AuCd$_4$-hP273 has $c = 5c_0$ and contains ~100 basic cells that gives ~2.73 atom/cell. This relates to the CuZn$_3$-*hP*3 phase with ~2.7 atoms/cell.

The case of FS-BZ design for the AuCd$_4$-$hP$273 phase just confirms the suggestion: "From the perspective of Jones theory, a metal structure is most stable if its Jones zone is as similar as possible to the free electron sphere comprised of the metals valence electrons" [10].

## 4. Conclusion

Phase sequence in the Cu–Zn alloy system consists of simple high-symmetry metallic structures *fcc*, *bcc* and *hcp* (α, β and ε) with an addition of two phases (γ-*cI*52 and δ-*hP*3) that are derivatives of the *bcc* structure. Effects of atomic site ordering appear upon temperature decrease at the transition from *bcc* to the CsCl-type structure (*cI*2 → *cP*2). Phases α, β and ε are completely site-disordered because Cu and Zn, the constituent elements of the alloy, are close neighbors in the periodic table and have minimal differences in the atomic size and electronegativity.

In the Au–Cd alloy system there are considerably larger number of the intermediate phases than in Cu–Zn. This is due to the substantial increase in effects of ordering and the contribution of the band structure energy of valence electrons. The structures of Au–Cd compounds can be combined into groups according to their relation to the high-symmetry structures: Au$_3$Cd-*tI*16 is related to *fcc*; Au$_2$Cd-*hP*4 and *hP*98 are related to *hcp*; martensitic phases AuCd-*mP*6, *oP*4 and *hP*18 are related to *bcc*. Au$_5$Cd$_8$-*cI*52 and AuCd$_3$-*hP*24 can be considered as structures derived from *bcc* with superlattices and vacancies. Several compounds have separate phase regions and are defined by the formation of the tetrahedral, icosahedral and trigonal-prismatic clusters such as in Au$_3$Cd$_5$-*tI*32, AuCd$_2$-*mC*72 and AuCd$_4$-*hP*273. The latter phase provides an example of a complex structure forming an almost completely spherical BZ polyhedron accommodating the FS.

By constructing the BZ-FS configurations for the Au–Cd compounds we can see the complex polyhedra with the BZ planes touching the FS that demonstrates significance of the band structure energy of valence electrons for the phase stability. Semi-quantitative characteristics of this contribution are the number of BZ planes close to FS and the estimation of how well the volume of FS fits the volume of BZ. The FS-BZ configurations are related to the physical properties of the alloy phases.

The driving force to form a variety of complex ordering superlattices in the Au–Cd alloys can be understood as an advantage of the band structure energy over the electrostatic energy. Stimulation of atomic site ordering for the Au and Cd atoms is provided by high difference in atomic numbers of constituent metals. Additional BZ planes that arise due to formation of the ordered supercells lead to formation of many-faced polyhedra that accommodate well the free-electron Fermi sphere.

## Acknowledgments

The authors gratefully acknowledge Dr. Olga Degtyareva for valuable discussion and comments. This work is supported by the Program "The Matter under High Energy Density" of the Russian Academy of Sciences.

**Table 1.** Structure parameters of several Au – Cd phases. Pearson symbol, space group and lattice parameters are from literature data. Fermi sphere radius $k_F$, ratios of $k_F$ to distances of Brillouin zone planes $\tfrac{1}{2} q_{hkl}$ and the filling degree of Brillouin zones by electron states $V_{FS}/V_{BZ}$ are calculated by the program BRIZ [18].

| Phase | Au$_2$Cd ht | Au$_2$Cd | AuCd ht | Au$_{1.1}$Cd$_{0.9}$ | AuCd | AuCd$_3$ |
|---|---|---|---|---|---|---|
| Pearson symbol | *hP*4 | *hP*98 | *cP*2 | *mP*6 | *oP*4 | *tP*2 |
| Structural data [a] | | | | | | |
| Space group | *P*6$_3$/*mmc* | *P*6$_3$/*mmc* | $Pm\bar{3}m$ | *P*12/*m*1 | *Pmma* | *P*4/*mmm* |
| lattice parameters (Å) | $a = 2.91$ $c = 9.558$ | $a = 20.433$ $c = 4.818$ | $a = 3.324$ | $a = 4.910$ $b = 3.089$ $c = 7.431$ $\beta = 105.38°$ | $a = 4.766$ $b = 3.151$ $c = 4.859$ | $a = 3.348$ $c = 2.873$ |
| FS – BZ data from the BRIZ program | | | | | | |
| z (number of valence electrons per atom) | 1.27 | 1.33 | 1.5 | 1.45 | 1.5 | 1.75 |
| $k_F$ (Å$^{-1}$) | 1.29 | 1.304 | 1.342 | 1.333 | 1.345 | 1.476 |
| Total number BZ planes | 32 | 26 | 12 | 18 | 16 | 12 |
| $k_F/(\tfrac{1}{2} q_{hkl})$ max min | 1.0348 0.915 | 1.060 0.929 | 1.004 | 1.0395 0.886 | 1.040 0.927 | 1.1125 1.1025 |
| $V_{FS}/V_{BZ}$ | 0.755 | 0.804 | 0.750 | 0.751 | 0.768 | 0.875 |

[a] Refs. [6,7]

**Table 2.** Structure parameters of several Au - Cd phases as given in the literature. The Fermi sphere radius $k_F$, the total number of Brillouin zone planes, the ratio of $k_F$ to distances of Brillouin zone planes ½ $q_{hkl}$ and the degree of filling of Brillouin zones by electron states $V_{FS}/V_{BZ}$ are calculated by the program BRIZ [18].

| Phase | Au$_3$Cd | AuCd | Au$_5$Cd$_8$ | Au$_3$Cd$_5$ | Au$_{36}$Cd$_{64}$ | AuCd$_3$ | AuCd$_4$ |
|---|---|---|---|---|---|---|---|
| Pearson symbol | *tI*16 | *hP*18 | *cI*52 | *tI*32 | *mC*72 | *hP*24 | *hP*273 |
| | | | Structural data | | | | |
| Space group | *I*4/*mmm* | *P*3 | *I*-43*m* | *I*4/*mcm* | *C*2/*m* | *P*6$_3$*cm* | *P*6$_3$/*m* |
| Lattice parameters (Å) | *a* = 4.116<br>*c* = 16.54<br>*c/a* =4.018 | *a* = 8.095<br>*c* = 5.794<br>*c/a* =0.716 | *a* = 10.03 | *a* = 10.728<br>*c* = 5.352<br>*c/a* = 0.499 | *a* = 14.735<br>*b* = 8.248<br>*c* = 12.714<br>*β* = 115.11 | *a* = 8.147<br>*c* = 8.511<br>*c/a* =1.045 | *a* = 21.326<br>*c* = 14.125<br>*c/a* =0.662 |
| Atomic volume (Å$^3$) | 17.51 | 18.27 | 19.404 | 19.25 | 19.54 | 20.38 | 20.38 |
| Reference | [7] | [7] | [7] | [7] | [33] | [7] | [36] |
| | | | FS – BZ data from the BRIZ program | | | | |
| *z* (number of valence electrons per atom) | 1.25 | 1.495 | 1.615 | 1.625 | 1.64 | 1.75 | 1.8 |
| $k_F$ (Å$^{-1}$) | 1.283 | 1.345 | 1.351 | 1.357 | 1.3567 | 1.3647 | 1.378 |
| Total number of BZ planes | 22 | 30 | 36 | 32 | 26 | 42 | 42 |
| *HKL*: 2$k_F$/$q_{hkl}$ | *105* : 1.053<br>*114* : 0.972<br>*200* : 0.844<br>*008* :0.840 | *121*: 1.032<br>*11-2* : 1.009<br>*300*: 1.0005 | *330*:1.017<br>*411*: 1.017 | *420*: 1.0362<br>*202*: 1.0283<br>*411*: 1.0108 | *-601*: 1.028<br>*132* : 1.0114<br>*-332*: 1.0203<br>*-424*: 1.0151<br>*-115*: 1.0151<br>*223* : 1.0182<br>*511*: 0.990 | *300*: 1.0216<br>*113*: 1.0116<br>*212*: 0.982 | *630* : 1.0206<br>*443* : 1.0175<br>*145* : 1.0147<br>*172* : 1.0139 |
| Filling of BZ with electron states $V_{FS}/V_{BZ}$ (%) | 77.0 | 89.7 | 93.3 | 95 | 88.5 | 90.5 | 96.2 |

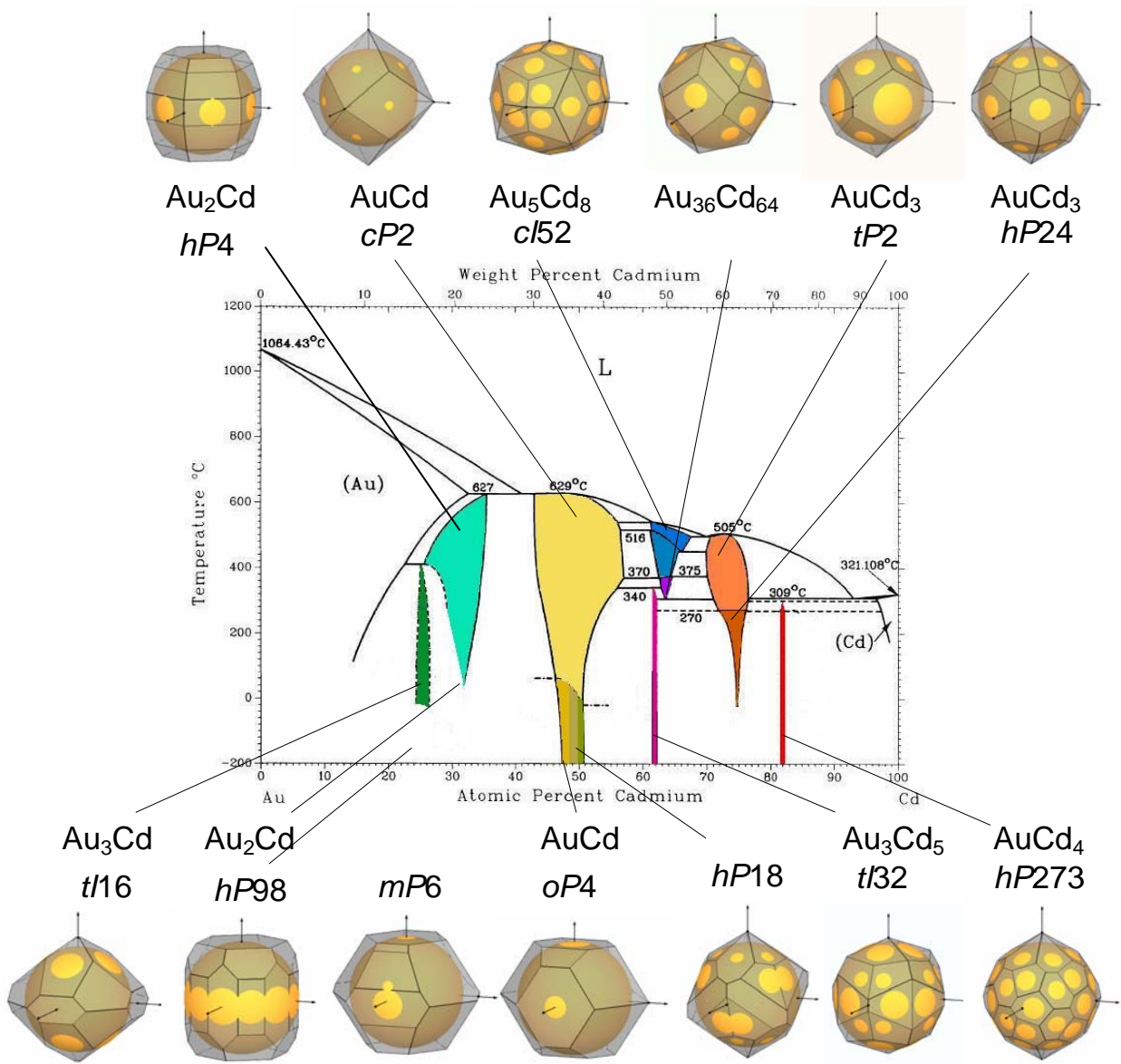

**Figure 1**. Au-Cd phase diagram adopted from Okamoto and Massalski [6]. Constructions of FS-BZ configurations are shown for the intermediate compounds with indicated compositions and structure types. Structural data and data from the BRIZ program are given in Tables 1 and 2.

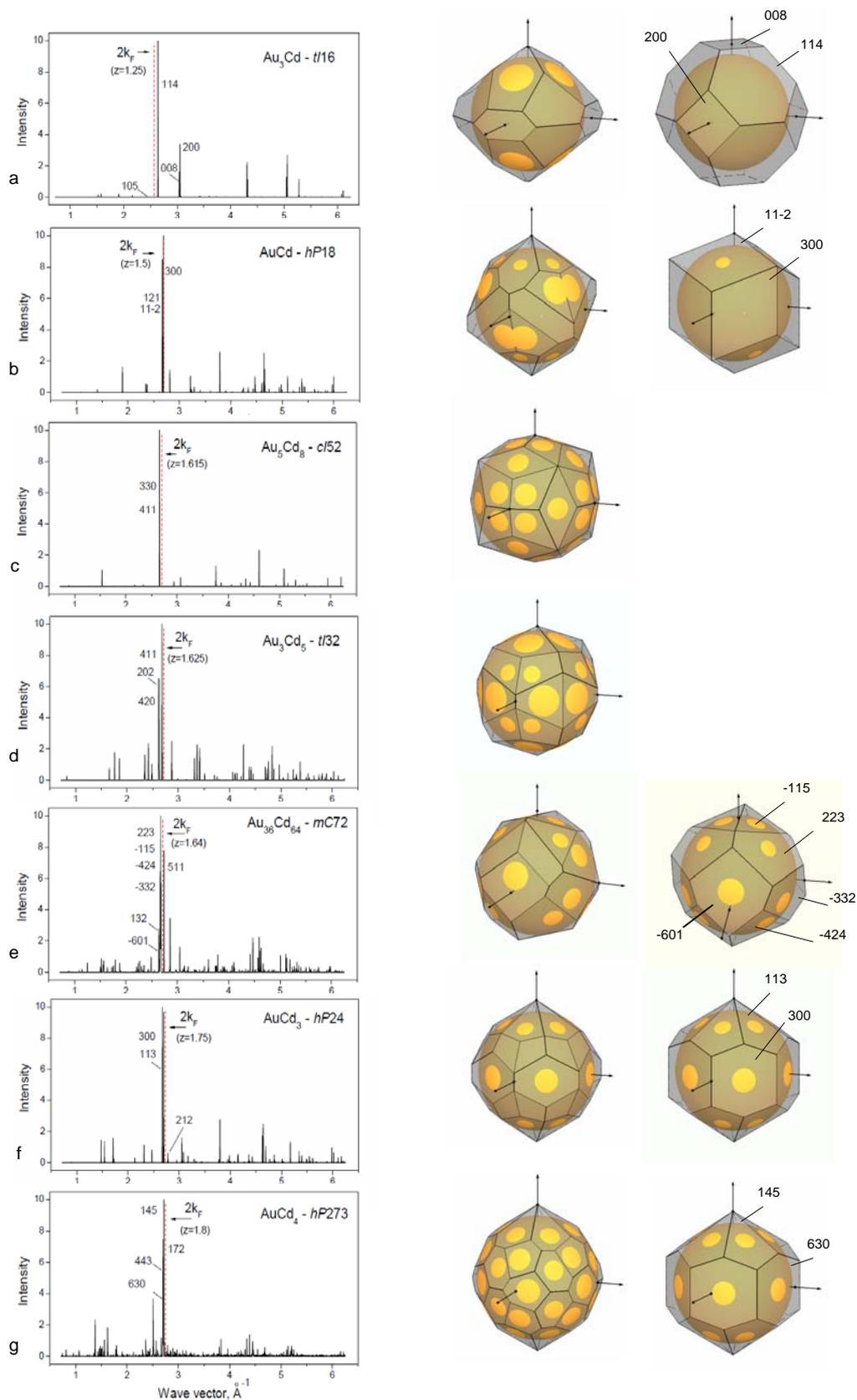

**Figure 2.** Calculated diffraction patterns for selected phases from Table 2 (left) and corresponding Brillouin-Jones zones with the inscribed Fermi spheres (middle). The position of $2k_F$ and the hkl indices of the planes used for the BZ construction are indicated on the diffraction patterns. Constructions of BZ planes corresponding to the strong reflections (right) show structural relationship to the basic cells (see text)